# Implication of the double-gating mode in hybrid photon counting detector for measurements of transient heat conduction in GaAs/AlAs superlattice structures

Authors

Denys Naumenko[a], Max Burian[a,b], Benedetta Marmiroli[a], Richard Haider[a], Andrea Radeticchio[a], Lucas Wagner[b], Luca Piazza[b], Lisa Glatt[b], Stefan Brandstetter[b], Simone Dal Zilio[c], Giorgio Biasiol[c] and Heinz Amenitsch[a]*

[a] Institute of Inorganic Chemistry, Graz University of Technology, Stremayrgasse 9/IV, Graz, 8010, Austria

[b] DECTRIS Ltd, Taefernweg 1, Baden-Daettwil, 5405, Switzerland

[c] CNR-IOM - Istituto Officina dei Materiali, SS 14, km 163.5, Basovizza (Trieste), 34149, Italy

Correspondence email: amenitsch@tugraz.at

**Synopsis** Use of double-gating mode implemented on modern hybrid photon counting system (EIGER2) helps to suppress influence of beam fluctuations in pump-probe experiments at synchrotron radiation facilities and provide better data quality.

**Abstract** Understanding and control of thermal transport in solids at nanoscale is crucial to engineer and to enhance properties of a new generation of optoelectronic, thermoelectric, and photonic devices. In this regard, semiconductor superlattice structures provide a unique platform to study phenomena associated with phonon propagations in solids such as heat conduction. Transient X-ray diffraction can directly probe atomic motions and therefore is among the rare techniques sensitive to phonon dynamics in a condensed matter. Here, we study optically induced transient heat conduction in GaAs/AlAs superlattice structures using EIGER2 detector. Benchmark experiments have been performed at the Austrian SAXS beamline at Elettra Sincrotrone Trieste that operated in the hybrid filling mode. We demonstrate that drifts of experimental conditions, such as synchrotron beam fluctuations, become less essential when utilizing EIGER2 double gating mode that results in a faster acquisition of high quality data and facilitates data analysis and data interpretation.

## 1. Introduction

Many fundamental physical, chemical, biochemical etc. processes in nature involve structural changes of matter on femtoseconds to picoseconds time scales that are the natural oscillation periods of atoms and molecules (Siders *et al.,* 1999, Ihee *et al.,* 2010). The visible lasers with fs/ps pulse widths are intensively used to optically pump and dynamically probe a wide range of atomic, molecular, solid-state, and plasma systems (Galli *et al.,* 1990, Ruello & Gusev, 2015, Fischer *et al.,* 2016, Basiri *et al.,* 2022). The drawback of all optical pump-probe techniques is that visible light cannot resolve atomic-scale features since it does not interact with the core electrons and nuclei that most directly indicate a structure. Nevertheless, it interacts with valence and free electrons and probes the ensuing dynamics inherently (Chergui & Collet, 2017, Bencivenga *et al.,* 2019, Maiuri *et al.,* 2020). Hard X-ray radiation, with wavelengths comparable with interatomic distances, is well suited to investigate structure and atomic rearrangement, and can measure structural dynamics in the interior of samples that are not transparent to ordinary light (Rose-Petruck *et al.,* 1999, Lindenberg *et al.,* 2017). Recently, high-brightness free electron lasers and synchrotron sources have been used in new classes of visible-pump / X-ray probe experiments to follow structural changes in solid-state systems, organic and protein crystals with excellent spatial resolution (Chergui & Collet, 2017, Bencivenga *et al.,* 2019, Maiuri *et al.,* 2020). Over the last years, the Austrian SAXS beamline at Elettra Sincrotrone Trieste has implemented such pump-probe X-ray diffraction/scattering setup and demonstrated that light-induced structural changes in semiconductor nanostructures and metal-organic frameworks on time scales ranging from picoseconds to seconds, respectively, (Burian *et al.,* 2020, Klokic *et al.,* 2022), can be successfully studied. Nevertheless, choosing the right data acquisition system in pump-probe experiments is always a challenge mainly because of intensity fluctuations and long term stability of light sources, beamline optics and sample environment. In this work an implication of double-gating mode of hybrid photon counting detector allows a better correction for instabilities of experimental conditions. We demonstrate this by decoupling complex phenomena at sub-nanosecond timescales in optically excited GaAs/AlAs superlattice (SL) structures.

## 2. Experimental Setup and Methodology

All experiments have been performed at the Austrian SAXS beamline at Elettra Sincrotrone Trieste that operated in the hybrid filling mode (Figure S1) as described in detail elsewhere (Burian *et al.,* 2020). The single bunch current (2 GeV electron energy) was set to 1 mA, resulting in X-ray pulse (8 keV energy) of approximately 100 ps full width at half-maximum (FWHM). Both the laser source (PHAROS, Light Conversion, Lithuania) and the X-ray detector (EIGER2, Dectris, Switzerland) have been synchronized to the storage ring time-base as depicted in Figure 1. A digital delay generator (P400, Highland-Electronics, USA) operated at one fourth of the ring clock frequency (289.4 kHz) and triggered the detector in order to select the single bunch. Temporal synchronization of the laser with the storage ring was achieved using the phase-comparison system (PhaseLock, TEM-Messtechnik, Germany) triggering the laser at one eighth of the ring clock frequency (144.7 kHz). In such configuration the laser repetition rate was set to a half of the detector one so that every second counted X-Ray probe pulse was synchronized with the laser pump pulse (Figure 1b) and the pumped image has been recorded (Counter A). The background unpumped image has been recorded using the Counter B shifted by a half period in the reference storage ring time frame. The frequency of trigger pulse train that is sent to the detector is doubled compared to the laser repetition rate. The laser wavelength was tuned to 515 nm (second harmonic of Yb:KGW laser, 1030 nm centre wavelength, 250 fs pulse width). The energy per pulse was set to 6 µJ while the laser beam diameter was set to 0.5 mm, resulting in the fluence of 3 mJ/cm$^2$ at the sample plane. Note the laser beam stabilization system (MRC, Germany) has been used to stabilize the position with an accuracy of 99.8 %. The X-Ray beam size has been set to 0.35 x 0.1 mm$^2$ (H x V) that results in almost square shape footprint in θ/2θ configuration at a scattering vector q = 22.2 ± 0.8 nm$^{-1}$ (Figure 2a). The details of spatial overlap of laser and X-Ray beams are given in (Burian *et al.*, 2020). The data integration is performed using the open software SAXSDOG for real-time azimuthal integration of 2D scattering images (Burian *et al.*, 2022). The data reduction and processing has been performed using IGOR Pro (IGOR Pro 7.0.8.1, WaveMetrics). The error propagation is calculated using the formalism described in (Burian *et al.*, 2020). The theoretical modelling of pump induced transient heat transfer has been performed using the udkm1Dsim toolbox (Schick *et al.*, 2014).

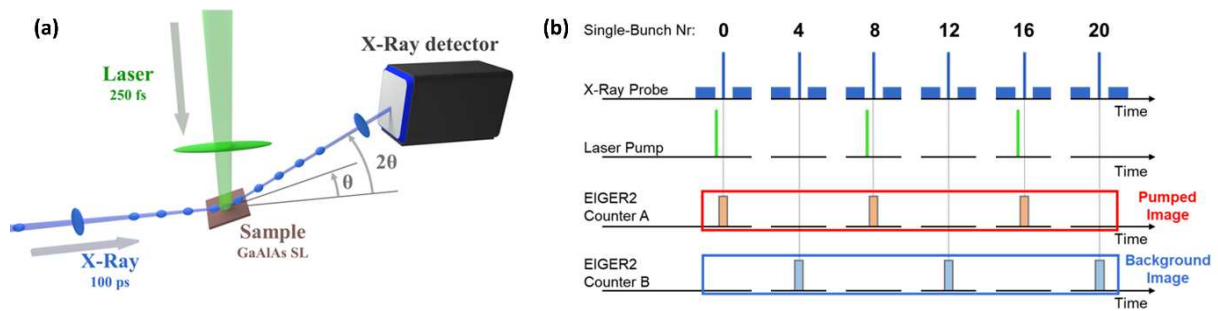

**Figure 1** Sketch of the experimental setup (a) and synchronization scheme of laser pump and X-ray probe to the storage ring filling pattern (b). Every 4$^{th}$ single bunch is displayed in panel (b). More detailed filling pattern is shown in Figure S1.

**3. Results**

**3.1. Sample Characterization**

The GaAs/AlAs SL structure has been grown on GaAs(001) substrate using the molecular beam epitaxy (MBE) technique. Such structures exhibit a small lattice mismatch (~0.2%) with interfacial mixing of <~3 monolayers (Robb & Craven, 2008, Cheaito *et al.*, 2018, Luckyanova *et al.*, 2018). The total thickness of the SL has been designed to be 535 nm, (4.5 nm GaAs + 6.2 nm AlAs) x 50 times. The periodicity and crystalline structure of SL have been verified by XRD and SEM as depicted in Figure 2 (a) and (b) respectively. A θ-2θ rocking-curve scan shows an outstanding agreement of the real superlattice structure with the simulations which have been performed using dynamical scattering theory (Schick *et al.*, 2014). The GaAs substrate diffraction peak at ambient temperature (300 K) has been identified at scattering vector $q$ equal to 22.23 nm$^{-1}$. It corresponds to (002) reflection of zinc blend GaAs crystal structure with the lattice constant of 0.5653 nm (Adachi, 1985). The intense peaks which arise from the secondary interference of diffracted X-rays on SL periodicity represent $0^{th}$, and $\pm 1^{st}$ SL diffraction orders as highlighted in Figure 2a. The evaluated SL period is 10.76 nm (4.53 nm of GaAs / 6.23 nm of AlAs) resulting in the total SL thickness of 538 nm that agrees with SEM measurements (Fig.2b). Prior to pump-probe experiments the SL has been thermalized. The downshift ($\Delta q = 0.8 \cdot 10^{-3}$ nm$^{-1}$) of SL $0^{th}$ order diffraction peak after laser induced thermalization implies that the averaged SL temperature has been raised to 306.7 K that has been further used as SL initial temperature for simulations of transient heat conduction (see Section 3.2).

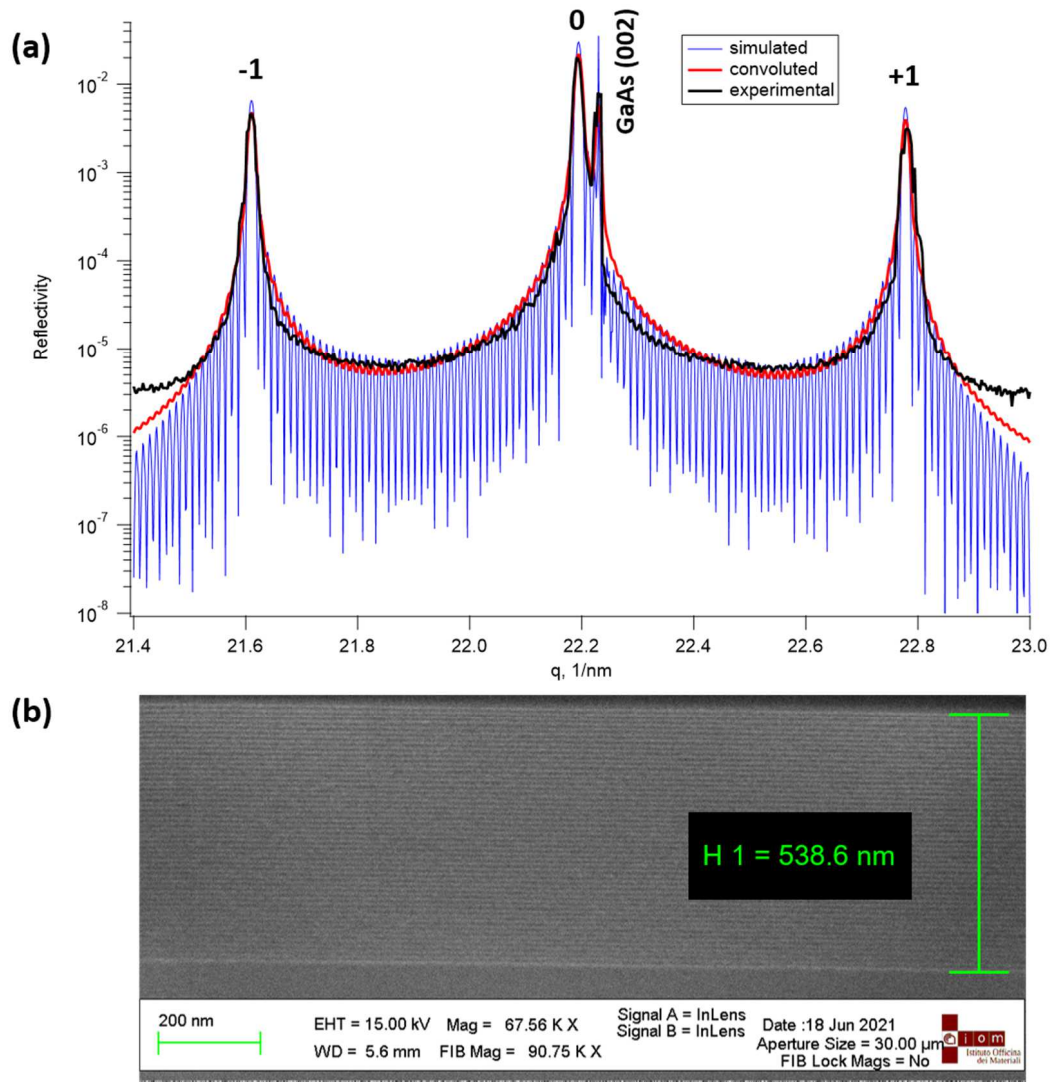

**Figure 2** (a) Simulated rocking curve of GaAs/AlAs SL (blue), its convolution with the instrumental function (red), and the experimental data (black). (b) Cross-sectional SEM image of corresponding SL with measured total SL thickness (tilt corrected).

**3.2. Double Gating Performance**

In order to obtain a temporal overlap of laser and single X-Ray pulses the transient signal has been measured at scattering vector of 22.19 nm$^{-1}$ that approximately corresponds to the half width at half maximum off-set of GaAs/AlAs SL 0$^{th}$ order peak at lower q (Figure 2a). In such conditions the largest signal contrast is expected due to the peak broadening and its shift towards lower q in consequence of pump induced heating of SL. The advantages of double gate mode can be clearly seen from Figure 3 where beam instabilities vanish simply dividing the signals or alternatively they can be subtracted.

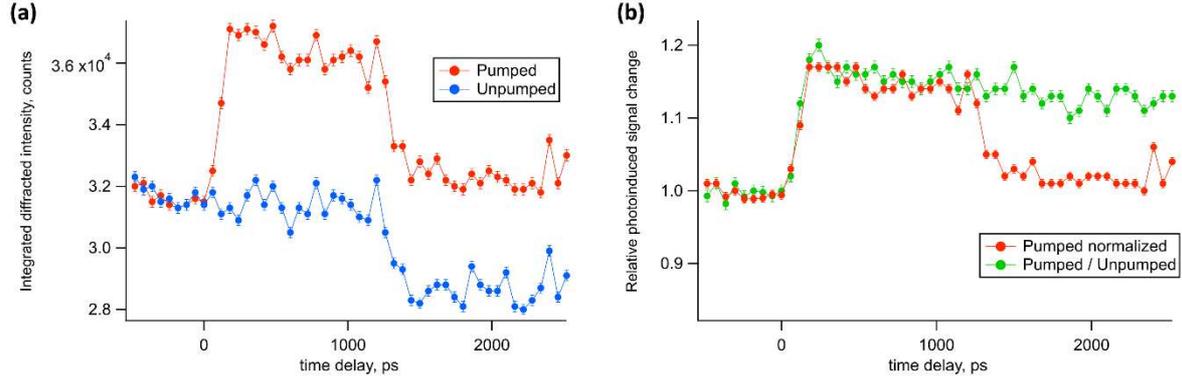

**Figure 3** (a) Integrated diffracted intensity of GaAs/AlAs SL $0^{th}$ order peak as function of delay time between the laser pump and X-Ray probe. (b) Relative photoinduced signal change in double (green) and single gate mode (red).

The absolute difference signals (|Counter A - Counter B|) measured across the $0^{th}$ and $\pm 1^{st}$ order SL peaks as function of delay time between the optical pump and X-Ray probe are displayed in Figure 4. The absolute values are taken in order to highlight that transient changes induced by the pump pulse dominantly occur due to the shift of corresponding GaAs/AlAs SL peaks rather than their broadening (Figure S2g). The downshift of all three SL peaks, $\Delta q = (7 \pm 0.5) \cdot 10^{-4}$ nm$^{-1}$, has been observed due to the transient SL temperature rise of $\Delta T = 5.7 \pm 0.4$ K that corresponds to 10 fm expansion of single unit cell constituting the SL.

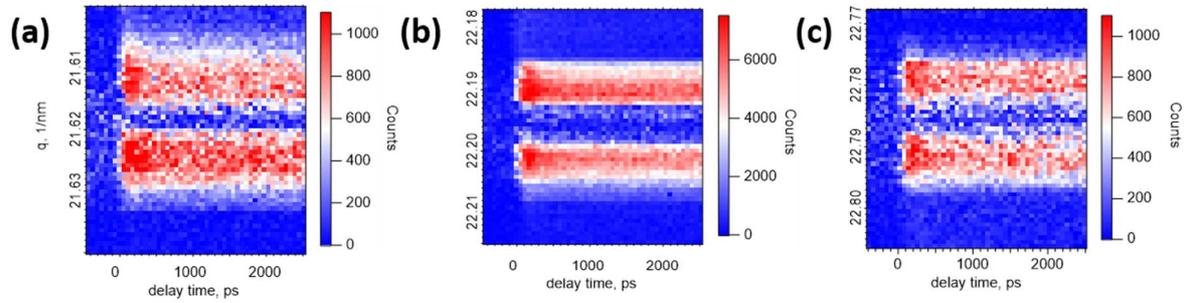

**Figure 4** Absolute difference of transient integrated intensity changes in the proximity of $-1^{st}$ (a), $0^{th}$ (b), and $+1^{st}$ (c) GaAs/AlAs SL peaks. The angular increment is 0.0005°, the time delay step is 60 ps.

For the further analysis and comparison with theoretical predictions we nevertheless use the normalized differential signal (Figure S2) that is given by:

$$S_{norm} = 2\frac{|A - B|}{A + B} \quad (1)$$

where A and B represent pumped and unpumped signals, Counter A and B, respectively. Such representation of transient signal for actual GaAs/AlAs SL (Figure 5) brings double benefit. First, the calculations of transient heat transfer at given temporal resolution can be performed for a single diffraction peak reducing computational time (Figure S3), and second, the further procedure of experimental data reduction for $0^{th}$ and $\pm 1^{st}$ SL peaks can be easily verified. This however is not valid in general. For instance, more sophisticated structures with additional buffer layers or SLs must be carefully examined.

### 3.3. Transient Heat Conduction & Simulations

Pump induced transient heat conduction has been modelled using the udkm1Dsim toolbox (Schick *et al.,* 2014). The udkm1Dsim is capable to simulate the structural dynamics in one-dimensional crystalline structures upon an ultrashort optical excitation. It therefore allows to evaluate cross-plane thermal conductivity, specific heat and group velocity. We have found that agreement between the experimental data on hundred ns time scale and simulations can only be obtained with significantly reduced thermal conductivity of GaAs/AlAs SL (Figure S4). This is not surprising since the reduction in cross plane thermal conductivity is well studied phenomenon in GaAs/AlAs SL structures (Luckyanova *et al.,* 2013, Cheaito *et al.,* 2018, Giri *et al.,* 2018). Thermal transport processes at nanoscale are complicated by the breakdown of the Fourier law of diffusion when size effects affect the mean free path of the heat carriers, which in semiconductor and dielectric materials are represented by acoustic phonons (Cahill *et al.,* 2014, Naumenko *et al.,* 2019). Different models that go beyond the diffusive regime, e.g., ballistic heat transport, have to be considered in order to describe such a phenomenology (Chen, 2021). Most experimental observations invoke the Casimir picture, wherein phonons travel ballistically or quasiballistically through the internal region of the sample and scatter diffusively at interfaces and boundaries (Luckyanova *et al.,* 2012, Chen, 2021). In this classical size regime, the phase information carried by phonons can be lost through the diffuse scattering at boundaries and by internal scattering processes. However, if the phase information is kept, it should be possible to control the conduction of heat by manipulating phonon waves through interference effects in periodic structures and thermal bandgap formation (Maldovan, 2015). The latter is possible due to a modification of the phonon dispersion relations which control the propagation of phonons by establishing densities of modes and group velocities. In the Casimir-Knudsen regime, the local heat flux is no longer proportional to the local temperature gradient as Fourier's law predicts but depends on the spatial temperature distribution (Chen, 2021). The udkm1Dsim toolbox (Schick *et al.,* 2014) provides such spatial temperature distribution which is an alternative to solving the Boltzmann transport equations or Monte Carlo simulations with highly uncertain boundary conditions (Zeng *et al.,* 2015, Anufriev *et al.,* 2017, Cheaito *et al.,* 2018, Beardo *et al.,* 2021, Chen, 2021).

Figure 5 shows normalized differential GaAs/AlAs SL signal as function of delay time between optical pump and X-ray probe. As simulations predict (Figure S3), all three experimental curves, $0^{th}$ order and two satellite SL peaks, are in a good agreement. The best signal-to-noise ratio at the same experimental conditions is observed for $0^{th}$ order GaAs/AlAs SL peak due to its higher diffraction efficiency. Nevertheless, the averaged signal is used for a comparison with simulations. As for long time scale experiment (Figure S4), the theoretical predictions based on bulk material properties fail to describe a transient GaAs/AlAs SL response at short (< 3 ns) time scale. This is especially evident during a temperature rise and initial strain release that occur at time scale shorter than 600 ps. At this stage the model gives a decent match only with reduced specific heat and group velocity. The heat conduction with reduced thermal conductivity well describes further dynamics that is valid until the SL is fully relaxed to its initial state. The adjusted GaAs/AlAs SL material properties employed for simulations and their relative decrease are summarized in Table S1. Based on a formalism of the kinetic theory (Appendix A) we also conclude that the phonon mean free path is reduced by 56%. Our current findings suggest that ultrafast experiments at sub-picosecond/picosecond time scale have to be performed in order to complete a full picture of dynamical response of optically excited GaAs/AlAs SL structures. However, for time scales above 100 ps the pump-probe setup implemented at the Austrian SAXS beamline at Elettra Sincrotrone Trieste demonstrates a high potential, which will be enhanced by reducing the X-ray pulse width with the Elettra 2.0 upgrade (Franciosi & Kiskinova, 2023).

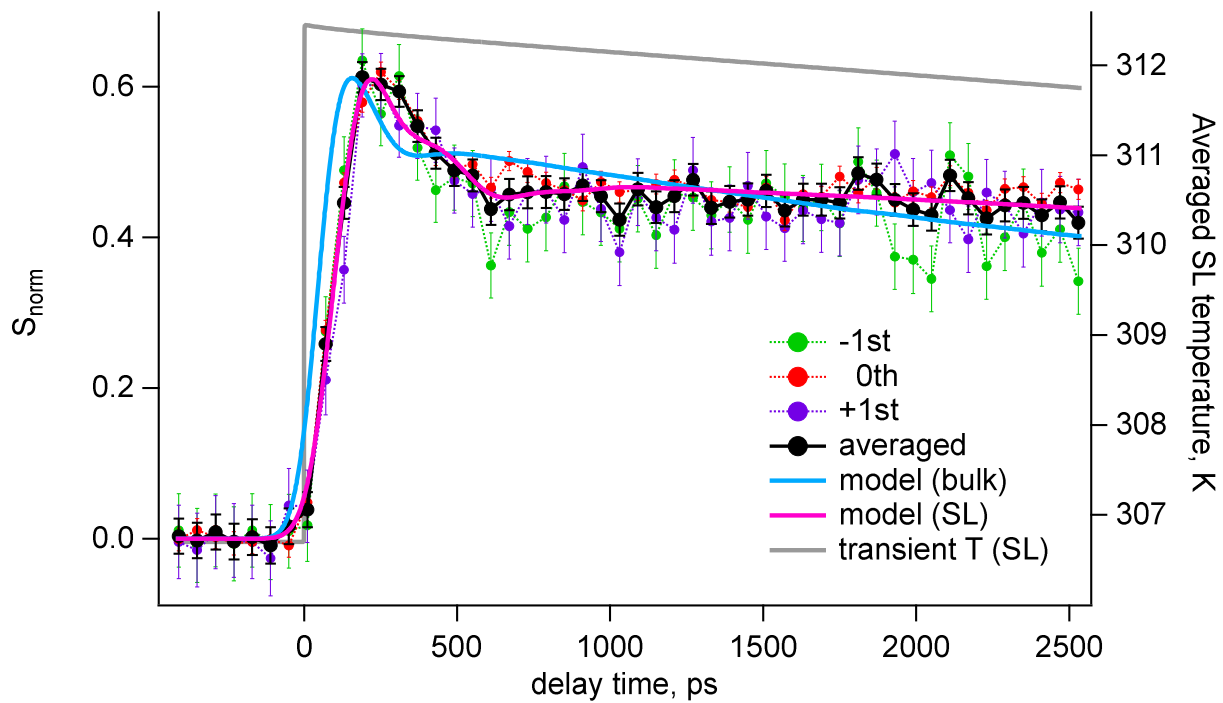

**Figure 5** Normalized transient differential signal for $-1^{st}$ (green), $0^{th}$ (red), and $+1^{st}$ (violet) GaAs/AlAs SL peaks. Their averaged trace is depicted in black. Blue and magenta solid lines

represent the SL response modelled using bulk material properties and adjusted ones, respectively. Grey solid line represents the average temperature of whole SL as function of delay time between optical pump and X-ray probe. The displayed curves have been obtained averaging both the experimental and theoretical data in the same angular range of 0.003° as displayed in Figure S2c.

## 4. Conclusion

Thermal transport processes at nanoscale are complicated by the breakdown of the Fourier law of diffusion. This is mostly due to classical size effects that lead to a significant reduction of thermal conductivity. Here, we demonstrated that in optically pumped 538 nm thick GaAs/AlAs superlattice structure the heat conduction with reduced thermal conductivity well describes a structural dynamic in nanosecond to microsecond time window. However, at short time scales (< 600 ps) both reduced specific heat and group velocity determine the rise and immediate response of SL normalized transient differential signal. These findings have been achieved using the upgraded version of picosecond pump-probe X-ray diffraction/scattering setup at the Austrian SAXS beamline at Elettra-Sincrotrone Trieste. We have shown that use of the double gating mode implemented in hybrid photon counting detector (EIGER2) helps supressing influence of beam fluctuations and provide better data quality. The laser feedback system on the other hand kept a laser beam trajectory with 99.8% accuracy avoiding sample temperature fluctuations. The best match between experimental and modelled data is obtained with 50% to 65% reduced specific heat, group velocity, and phonon mean free path that resulted in 12.5-fold decrease of thermal conductivity. Our findings suggest that better temporal resolution at shorter time scale (< 100 ps) is required in order to complete a full picture of dynamical response of optically excited GaAs/AlAs SL structures. Nevertheless, for time scales above 100 ps the pump-probe setup at the Austrian SAXS beamline demonstrates its high potential and a new class of experiments on bio-structures in liquid, photo-switchable systems, T-jump experiments in water etc. is foreseen in the nearest future.

**Acknowledgements**     The authors are grateful to the CERIC-ERIC consortium for providing access to the TUG's SAXS beamline at ELETTRA – Sincrotrone Trieste. Authors from DECTRIS Ltd. declare a conflict of interest, as DECTRIS Ltd. is the developer and manufacturer of the mentioned EIGER2 detector.

**References**

Adachi, S. (1985). *Journal of Applied Physics, 58*(3).

Anufriev, R., Ramiere, A., Maire, J., & Nomura, M. (2017). *Nature Communications*, *8*, 1–8.

Basiri, A., Rafique, M. Z. E., Bai, J., Choi, S., & Yao, Y. (2022). *Light: Science and Applications*, *11*(1), 102.

Beardo, A., Knobloch, J. L., Sendra, L., Bafaluy, J., Frazer, T. D., Chao, W., Hernandez-Charpak, J. N., Kapteyn, H. C., Abad, B., Murnane, M. M., Alvarez, F. X., & Camacho, J. (2021). *ACS Nano*, *15*(8), 13019–13030.

Bencivenga, F., Mincigrucci, R., Capotondi, F., Foglia, L., Naumenko, D., Maznev, A. A., Pedersoli, E., Simoncig, A., Caporaletti, F., Chiloyan, V., Cucini, R., Dallari, F., Duncan, R. A., Frazer, T. D., Gaio, G., Gessini, A., Giannessi, L., Huberman, S., Kapteyn, H., Masciovecchio, C. (2019). *Science Advances*, *5*(7), eaaw5805.

Burian, M., Marmiroli, B., Radeticchio, A., Morello, C., Naumenko, D., Biasiol, G., & Amenitsch, H. (2020). *Journal of Synchrotron Radiation*, *27*, 51–59.

Burian, M., Meisenbichler, C., Naumenko, D., & Amenitsch, H. (2022). *Journal of Applied Crystallography*, *55*, 677–685.

Cahill, D. G., Braun, P. V., Chen, G., Clarke, D. R., Fan, S., Goodson, K. E., Keblinski, P., King, W. P., Mahan, G. D., Majumdar, A., Maris, H. J., Phillpot, S. R., Pop, E., & Shi, L. (2014). *Applied Physics Reviews*, *1*(1).

Cheaito, R., Polanco, C. A., Addamane, S., Zhang, J., Ghosh, A. W., Balakrishnan, G., & Hopkins, P. E. (2018). *Physical Review B*, *97*(8), 1–7.

Chen, G. (2021). *Nature Reviews Physics*. 3(8), 555–569

Chergui, M., & Collet, E. (2017). *Chemical Reviews*, *117*(16), 11025–11065.

Durbin, S. M., Clevenger, T., Graber, T. & Henning, R. (2012). Nat. Photon. 6, 111–114.

Fischer, M. C., Wilson, J. W., Robles, F. E., & Warren, W. S. (2016). *Review of Scientific Instruments*, *87*(3), 031101.

Franciosi, A., Kiskinova, M., (2023). *Eur. Phys. J. Plus*, *138, 79*, https://doi.org/10.1140/epjp/s13360-023-03654-6.

Galli, G., Martin, R. M., Car, R., & Parrinello, M. (1990). *Science*, *250*(4987), 1547–1549.

Ihee, H., Wulff, M., Kim, J., & Adachi, S. ichi. (2010). *International Reviews in Physical Chemistry*, *29*(3), 453–520.

Giri, A., Braun, J. L., Shima, D. M., Addamane, S., Balakrishnan, G., & Hopkins, P. E. (2018). *Journal of Physical Chemistry C*, *122*(51), 29577–29585.

Klokic, S., Naumenko, D., Marmiroli, B., Carraro, F., Linares-Moreau, M., Zilio, S. D., Birarda, G., Kargl, R., Falcaro, P., & Amenitsch, H. (2022). *Chemical Science*, *13*(40), 11869–11877.

Lindenberg, A. M., Johnson, S. L., & Reis, D. A. (2017). *Annual Review of Materials Research*, *47*, 425–449.

Luckyanova, M. N., Garg, J., Esfarjani, K., Jandl, A., Bulsara, M. T., Schmidt, A. J., Minnich, A. J., Chen, S., Dresselhaus, M. S., Ren, Z., Fitzgerald, E. a, & Chen, G. (2012). *Science*, *338*, 936–939.

Luckyanova, M. N., Johnson, J. A., Maznev, A. A., Garg, J., Jandl, A., Bulsara, M. T., Fitzgerald, E. A., Nelson, K. A., & Chen, G. (2013). *Nano Letters*, *13*(9), 3973–3977.

Luckyanova, M. N., Mendoza, J., Lu, H., Song, B., Huang, S., Zhou, J., Li, M., Dong, Y., Zhou, H., Garlow, J., Wu, L., Kirby, B. J., Grutter, A. J., Puretzky, A. A., Zhu, Y., Dresselhaus, M. S., Gossard, A., & Chen, G. (2018). *Science Advances,* 4(12), aat9460.

Maiuri, M., Garavelli, M., & Cerullo, G. (2020). *Journal of the American Chemical Society*, *142*(1), 3–15.

Maldovan, M. (2015). *Nature Materials*, *14*(7), 667–674.

Naumenko, D., Mincigrucci, R., Altissimo, M., Foglia, L., Gessini, A., Kurdi, G., Nikolov, I., Pedersoli, E., Principi, E., Simoncig, A., Kiskinova, M., Masciovecchio, C., Capotondi, F., & Bencivenga, F. (2019). *ACS Applied Nano Materials*, *2*(8), 5132–5139.

Robb, P. D., & Craven, A. J. (2008). *Ultramicroscopy*, *109*(1), 61–69.

Rose-Petruck, C., Jimenez, R., Guo, T., Cavalleri, A., Siders, C. W., Ráksi, F., Squier, J. A., Walker, B. C., Wilson, K. R., & Barty, C. P. J. (1999). *Nature*, *398*(6725), 310–312.

Ruello, P., & Gusev, V. E. (2015). *Ultrasonics*, *56*, 21–35.

Schick, D., Bojahr, A., Herzog, M., Shayduk, R., Von Korff Schmising, C., & Bargheer, M. (2014). *Computer Physics Communications*, *185*(2), 651–660.

Siders, C. W., Cavalleri, A., Sokolowski-Tinten, K., Tóth, C., Guo, T., Kammler, M., Horn Von Hoegen, M., Wilson, K. R., Von Der Linde, D., & Barty, C. P. J. (1999). *Science*, *286*(5443), 1340–1342.

Zeng, L., Collins, K. C., Hu, Y., Luckyanova, M. N., Maznev, A. A., Huberman, S., Chiloyan, V., Zhou, J., Huang, X., Nelson, K. A., & Chen, G. (2015). *Scientific Reports*, *5*, 1–10.

**Appendix A.**

A dependence of thermal conductivity k on phonon properties is given by (Chen, 2021):

$$k = \frac{1}{3}\int_0^{\omega_{max}} C(\omega)v(\omega)\Lambda(\omega)d\omega \quad (2)$$

Where ω is the phonon angular frequency, C(ω) is the spectral volumetric specific heat, v(ω) is the group velocity, and Λ(ω) is the phonon mean free path. Using the relative values $\frac{\text{GaAs/AlAs SL}}{\text{GaAs/AlAs}^*}$ (Table S1), i.e. the ratio of adjusted GaAs/AlAs SL material properties to volume weighted average values, one can obtain:

$$\frac{k}{k^*} = \frac{Cv\Lambda}{C^*v^*\Lambda^*}; \quad \frac{\Lambda}{\Lambda^*} = 0.44$$

i.e. 56% decrease in the phonon mean free path.

# Supporting information

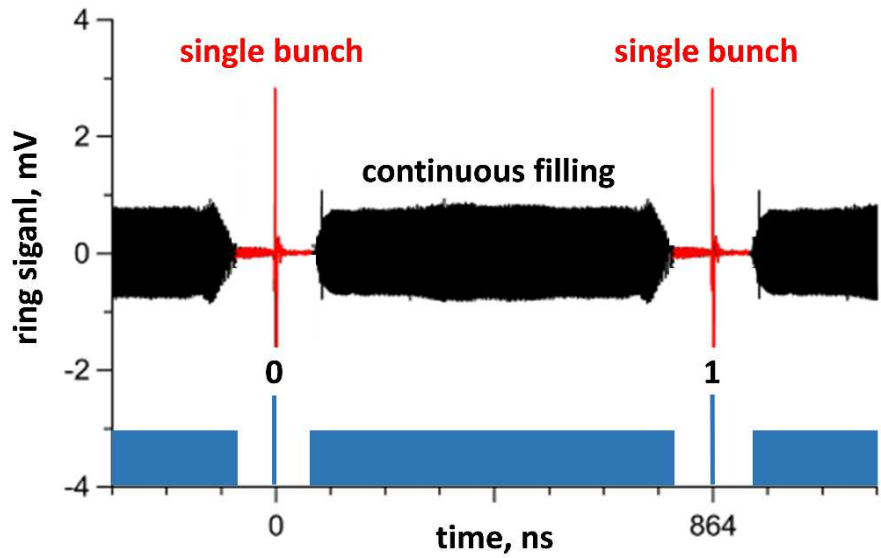

**Figure S1** Ring signal of the Elettra storage ring in hybrid filling mode, consisting of continuous filling regime (black), a dark gap and a single bunch in the dark gap centre (both in red). The bottom part (blue) schematically represents a hybrid filling mode where single bunches are sequentially numbered.

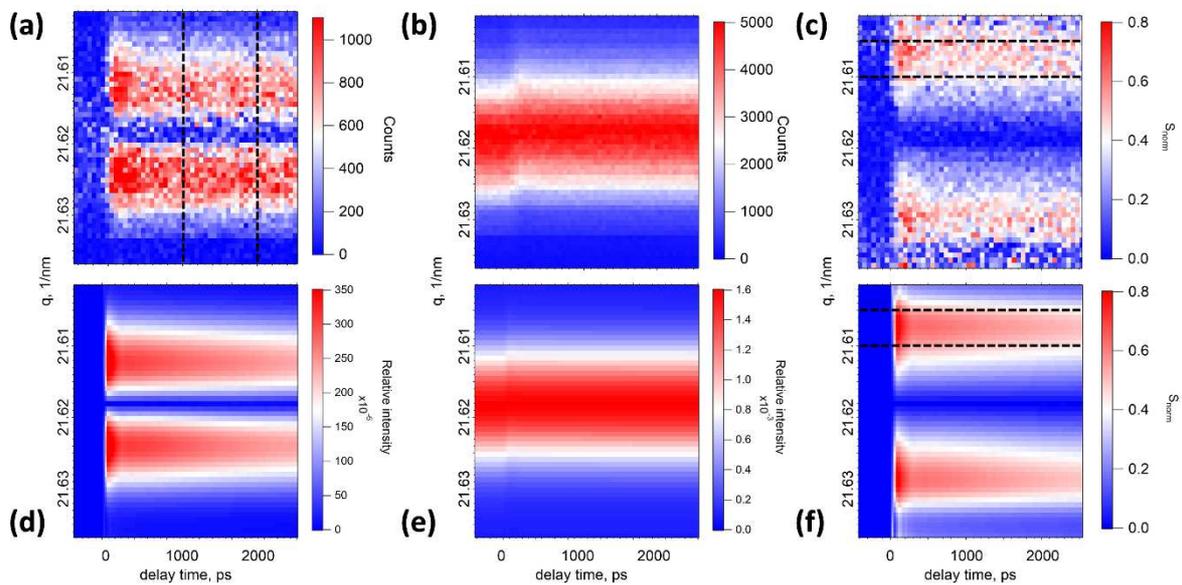

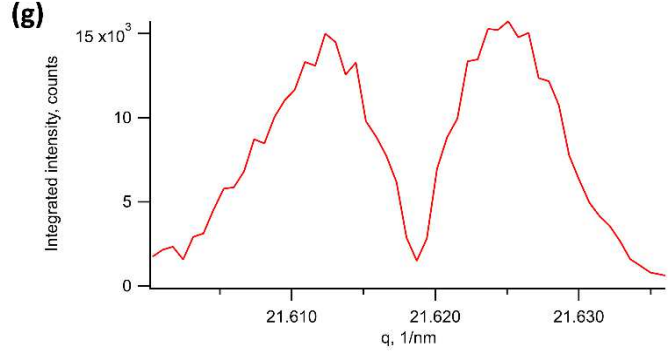

**Figure S2** Experimental (a-c) and simulated (d-f) transient signals in the proximity of -1$^{st}$ SL peak. Panels (a, d) represent absolute difference signal; the sum is depicted in panels (b, e) while panels (c, f) show normalized differential signal given by Eq. 1 in the main text. The angular increment is 0.0005°, the time delay step is 60 ps and 30 ps for experimental (a-c) and simulated (d-f) maps, respectively. The black vertical dashed lines in panel (a) represent time window from 1000 ps to 2000 ps in which the signal has been integrated and plotted in panel (g) as function of scattering vector $q$ testifying that the peak broadening is negligible. The black horizontal dashed lines in panels (c, f) represent the angular regions of 0.003° in which the signal ($S_{norm}$) has been averaged and plotted in Figure 5 in the main text.

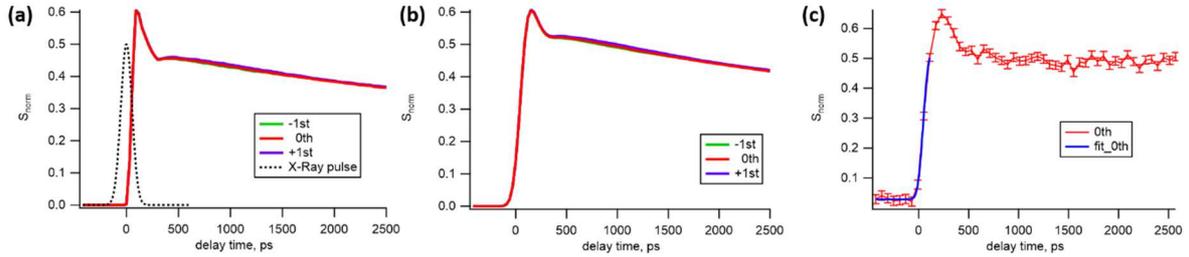

**Figure S3** Simulated (a) and convoluted (b) transient response of 0$^{th}$ and ±1$^{st}$ SL peaks as function of delay time between optical pump and X-Ray probe. The X-ray pulse width (104.6 ± 5.5 ps FWHM) has been determined from an error-function fit (Durbin *et al.*, 2012) of 0$^{th}$ order SL peak (c) and depicted with black dotted line in panel (a). Note that the fit range in (c) is determined by the 0$^{th}$ order peak position in panel (a), i.e. at 90 ps delay time.

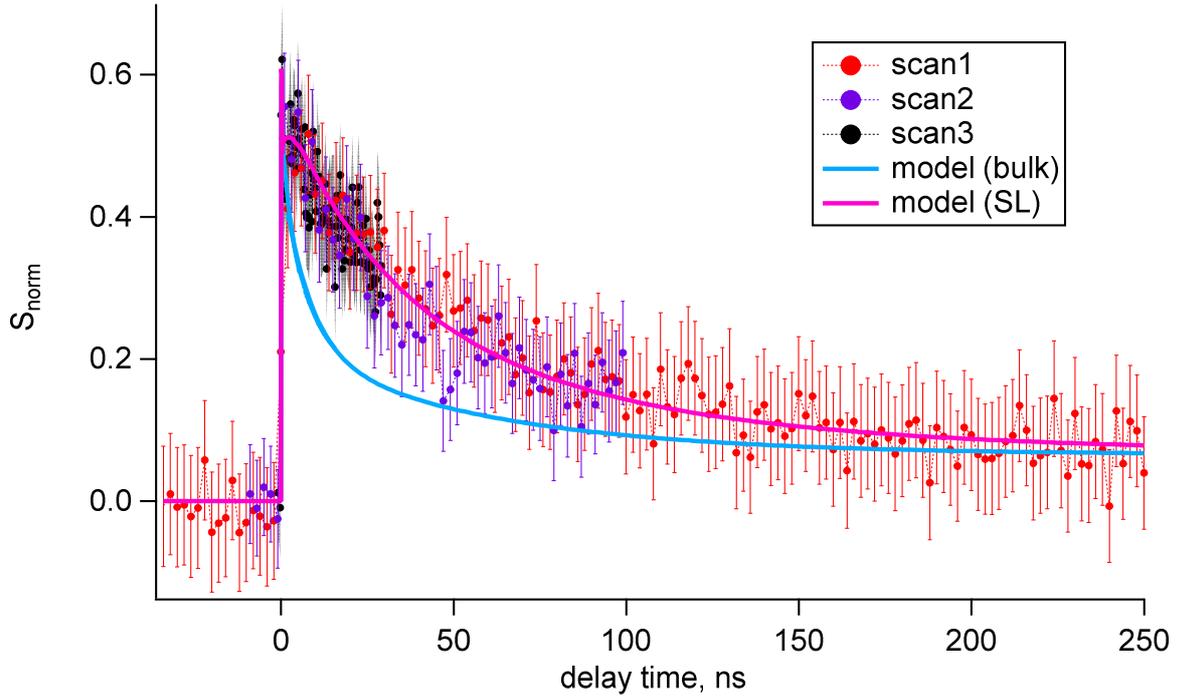

**Figure S4** Normalized differential transient signals for $0^{th}$ GaAs/AlAs SL peak measured in different time intervals with a variable time step (scan1-scan3). Blue and magenta solid lines represent the SL response modelled using bulk material properties and adjusted ones, respectively (see Table S1). Note that the thermal conductivity of 0.08 W/(m K) is in a good agreement with the data obtained by another methods (Luckyanova *et al.,* 2013, Cheaito *et al.,* 2018).

**Table S1** Bulk GaAs, AlAs, volume weighted average (GaAs/AlAs *) and adjusted GaAs/AlAs SL material properties employed for simulations as well as their relative changes $\frac{\text{GaAs/AlAs SL}}{\text{GaAs/AlAs}^*}$.

|  | GaAs[a] | AlAs[a] | GaAs/AlAs * | GaAs/AlAs SL | $\frac{\text{GaAs/AlAs SL}}{\text{GaAs/AlAs}^*}$ |
|---|---|---|---|---|---|
| Lattice constant, Å | 5.653 | 5.661 | 5.658 * | 5.658 * | 1 |
| Thermal expansion coefficient, (x$10^{-6}$ K) | 6.4 | 5.2 | 5.7 * | 5.7 * | 1 |
| Specific heat, J/(kg K) | 330 | 450 | 400 | 150 | 0.38 |
| Sound velocity, km/s | 4.73 | 5.65 | 5.26 | 2.50 | 0.48 |
| Thermal conductivity, W/(m K) | 55 | 91 | 76 | 6 | 0.08 |

[a] (Adachi, 1985).

* volume weighted average.